\documentstyle[12pt,a4,epsfig]{article}
\newcommand{\beq}[1]{\begin{equation}\label{#1}}
\newcommand{\eeq}{\end{equation}}
\newcommand{\beqar}[1]{\begin{eqnarray}\label{#1}}
\newcommand{\eeqar}{\end{eqnarray}}

\newcommand{\al}{\alpha}
\newcommand{\be}{\beta}  

\newcommand{\ga}{\gamma}
\newcommand{\de}{\delta}
\newcommand{\De}{\Delta}

\newcommand{\ka}{{\kappa}}
\newcommand{\la}{\lambda}

\newcommand{\si}{\sigma}

\newcommand{\Ga}{\Gamma}
\newcommand{\om}{\omega}

\def\eq#1{{Eq.~(\ref{#1})}}
\begin{document}
\vspace*{-2cm}
\begin{flushright}
 NTZ 15/89\\
\end{flushright}
\vspace{2cm}
\begin{center}
{\LARGE \bf 
 Small-$x$ behaviour of the polarized \\ photon  structure function $F_3^\ga(x,Q^2)$ 
\footnote{Supported by 
German Bundesministerium f\"ur Bildung, Wissenschaft, Forschung und
Technologie, grant No. 05 7LP91 P0, and 
by Volkswagen Stiftung}}\\[2mm]
\vspace{1cm}
{\bf B. Ermolaev}$^{\dagger \$}$, {\bf R.~Kirschner}$^\dagger$,
{\bf and L.~Szymanowski}$^{\dagger \#}$ \\
\vspace{1cm}

$^\dagger$Naturwissenschaftlich-Theoretisches Zentrum  \\
und Institut f\"ur Theoretische Physik, Universit\"at Leipzig
\\ 
Augustusplatz 10, D-04109 Leipzig, Germany
\\ 
\vspace{2em}
$^{\$}$ Ioffe Physico-Technical Institute, St.~Petersburg 194021, Russia \\

\vspace{2em}
$^{\#}$
Soltan Institut for Nuclear Studies, Ho\.za 69, 00-681 Warsaw, Poland

\end{center}   

\vspace{1cm}
\noindent{\bf Abstract:}
We study the small-$x$ behaviour of the polarized photon structure function
$F_3^{\ga}$, measuring the gluon transversity distribution, in the leading
logarithmic approximation of perturbative QCD.
There are two contributions, both arising from two-gluon exchange.
The leading contribution to small-$x$ is related to the BFKL pomeron and
behaves like $x^{-1-\om_2}$, $\om_2 = {\cal O}(\al_S)$.
The other contribution includes in particular the ones summed by
the DGLAP equation and behaves like $x^{1-\om_0^{(+)}}$, $\om_0^{(+)} =
{\cal O}(\sqrt{\al_S})$.

\vspace*{\fill}
\eject
\newpage

\section{Introduction}
\setcounter{equation}{0}

The deep-inelastic scattering off a  photon attracted much attention over 
20 years. The idea of extracting the direct contribution to the 
structure function, which is calculable from theory without additional 
assumptions \cite{Witten}, motivated experimental studies more involved
compared to the standard lepton-nucleon reactions (\cite{photon} and
references therein).

The extension of the deep-inelastic photon cross section on a target of spin
1 involves structure functions which have no analogon in the
usual spin $\frac{1}{2}$ case. The Lorentz structure of photon-photon
amplitudes have been studied by many authors, e.g. \cite{f-loop}.
The expression for the deep-inelastic cross section on general
spin 1 target is given in \cite{HJM}. We recall the expression for 
the photon case
\beqar{decomposition}
&&\frac{d^2\,\si}{dq^2\, dx} \sim L^{\mu \nu}\,W_{\mu \nu}\;\;,
\nonumber \\
&&W^{\mu \nu} = \left(F_1 - \frac{1}{2}F_3^{\ga}  \right)\left( 
g^{\mu \nu} - \frac{q^\mu q^\mu}{q^2}  \right)
\epsilon_\perp^{*\rho}\epsilon_{\perp \rho} \nonumber \\
&& - \left(F_2 - \frac{1}{2}F_3^{\ga}  \right)
\left( p^\mu - \frac{pq}{q^2}q^\mu \right)
\left( p^\nu - \frac{pq}{q^2}q^\nu \right)
\epsilon_\perp^{*\rho}\epsilon_{\perp \rho} \nonumber \\
&&+ \frac{1}{2}F_3^{\ga} \left( \epsilon_\perp^{*\mu}\epsilon_\perp^{\nu}
+ \epsilon_\perp^{*\nu}\epsilon_\perp^{\mu} \right) 
+ i\,g_1\;\epsilon^{\mu \nu \la \al}q_\la\, S_\al \;\;. 
\eeqar 
Here $\epsilon_\perp^{\mu}$ is the transverse (with respect to $p$ and $q$)
part of the target photon helicity vector. $ S_\al$ is the spin vector
given by $S_\al = i\,\epsilon^{\al \be \ga \tau}
\epsilon_{\perp \be}\epsilon_{\perp \ga}^*p_\tau\;$.

Comparing with the proton target case we see that indeed $F_3^{\ga}$
is the interesting new structure function.
It can be measured if the target photon is polarized; the polarization of the
virtual photon, i.e. of the lepton scattered with large momentum transfer,
is not necessary.
This structure function measures the parton polarization in an essentially 
different way compared to $g_1$. It is related to the virtual photon - photon
scattering amplitude with helicity flip \cite{HJM}, \cite{J-M89}. In order
to give this amplitude the probability interpretation of a parton 
distribution one has to change the basis from helicity (circular polarization)
to linear polarizations. $F_3^{\ga}$ measures the asymmetry of linear
polarizations, the transversity of gluons.

$F_3^{\ga}$ is similar to the structure function $h_1$, which appears in
the Drell-Yan cross-section and measures the transversity of quarks
\cite{transversity}. We show that there is a far-reaching analogy of 
the small-$x$ behaviour of both $F_3^{\ga}$ and $h_1$. The small-$x$
behaviour  of $h_1$ has been analyzed recently \cite{h1}.
There are recent results on the $Q^2$ evolution of $h_1$ \cite{h1Q}.
Related questions have been discussed in the workshop on spin physics at
HERA \cite{HERA}.

In the present paper we investigate the polarized photon structure function
$F_3^{\ga}$ 
at small-$x$. In this asymptotics it is appropriate to consider the
$t$-channel exchange of the corresponding forward scattering amplitude
and to study the leading singularities on the complex angular
momentum plane of the corresponding $t$-channel partial wave.

In general, $P$ parity and a minimal angular momentum are the relevant
$t$-channel quantum numbers for polarized structure functions.
The unpolarized parton distributions are related to positive
parity exchange, the helicity distributions to negative parity exchange.
The second type of polarized parton distributions, the transverse
polarization asymmetries, are related to the exchange of longitudinal
angular momentum 1 (quark transversity) or 2 (gluon transversity).

It is convenient to study the perturbative Regge asymptotics in the
framework of the high-energy effective action \cite{eff-action}, \cite{LKSsym},
 leading in particular to reggeized gluons and quarks.
Two leading reggeized gluons interacting via effective vertices result in 
the BFKL pomeron. Non-leading gluonic reggeons have been studied recently
\cite{gr}. The latter are relevant in particular for polarized structure 
functions since they are able to to transfer odd parity or non-vanishing 
longitudinal angular momentum projection ($s$-channel helicity) $\si$.
The leading gluonic reggeon corresponds to $\si =0$ and the leading 
reggeized quark to $\si =\frac{1}{2}$. The multiple  exchange of reggeons
gives rise to the tree-level energy behaviour $s^{\alpha_0}$, where
\beq{a}
\alpha_0 = 1 - \sum \si_i \;\;. 
\eeq
\noindent This can be considered as the extension of
the known Asimov rule \cite{Asimov} to non-leading exchanges.
Helicity 2 exchange by two gluons, relevant for our case, can occur
in two ways:

(1) by non-leading reggeons ($\si_1=0$, $\si_2=2$),  
$\alpha_0 = -1$

(2) by leading reggeons ($\si_1=\si_2=0$) with a non-vanishing
longitudinal projection of orbital momentum, $\alpha_0 = +1$.

 We shall encounter contributions of both cases.

After recalling in sec. 2 the known results about the photon spin-flip
amplitude \cite{f-loop}, \cite{Zima}, \cite{HJM}, \cite{J-M89},
about the one-loop coefficient function and the parton splitting function
\cite{Schmidt}, \cite{A-M}, \cite{Bukhvostov}, we sum in sec. 3 the 
double-logarithmic corrections. This concerns a contribution to
$F_3^{\ga}$ behaving like $x^{1-\om_0^{(+)}}$, corresponding to
the case (1) above with $\alpha_0 = -1$. We calculate the partial wave
 whose singularity position gives
 rise
to this behaviour and to the related resummed anomalous dimensions near
angular momentum $j=-1$.

The contribution of the BFKL pomeron to the photon spin-flip amplitude
is known \cite{BL-borel}. In sec.~4 we recall the impact factors relevant for the 
photon structure function and the branch püoint singularity of the 
conformal spin or helicity $n=2$ term in the BFKL solution.
This contribution to $F_3^{\ga}$ behaves like $x^{-1-\om_2}$
corresponding to the case (2) above with $\alpha_0=+1$.

\section{Lowest order contribution and $Q^2$ evolution }
\setcounter{equation}{0}

The spin structure function of the photon $F_3^{\ga}$  
is given by the imaginary
part of the virtual photon-photon forward scattering amplitude 
with spin flip of
both scattered photons \cite{J-M89}
\beqar{2.1}
&&W^{\mu \nu;\al \be}\Bigg|_{\De=2} = \left(
P^{\mu \nu;\al \be}_{(+-;-+)} +
P^{\mu \nu;\al \be}_{(-+;+-)}\right) F_3^\ga(x,Q^2) \nonumber \\
&&
P^{\mu \nu;\al \be}_{(+-;-+)} +
P^{\mu \nu;\al \be}_{(-+;+-)} = \frac{1}{2}\left(
g^{\mu \al}_{\perp} g^{\nu \be}_{\perp} +  g^{\mu \be}_{\perp}
g^{\nu \al}_{\perp} - g^{\mu \nu}_{\perp} g^{\al \be}_{\perp} \right) \;\;.
\eeqar
$W^{\mu \nu}$ in (\ref{decomposition}) is related to (\ref{2.1})
by $W^{\mu \nu}= W^{\mu \nu;\al \be}\epsilon_{\perp \al} \epsilon_{\perp \be}^*$. 
The transverse subspace is the one orthogonal to the momentum $q$
($q^2=-Q^2$) of the virtual photon and $p$ ($p^2=0$)  of the photon target.

Consider first the lowest order contribution to this amplitude
arising from a fermion loop, Fig.~1. This has been calculated by many
authors long ago \cite{f-loop}, \cite{Zima}. We  use calculations by
Zima \cite{Zima} and pick up the projection \eq{2.1} from the result
\beq{2.2}
F_3^{(0)}(x,Q^2) = \frac{e^4}{4\pi^2}\,x^2\,N \;\;,
\eeq
where $e$ is the electromagnetic coupling constant and $N$ denotes the number 
of
colours.

\begin{figure}[htb]
\label{ffig1}
\begin{center}  
\epsfig{file=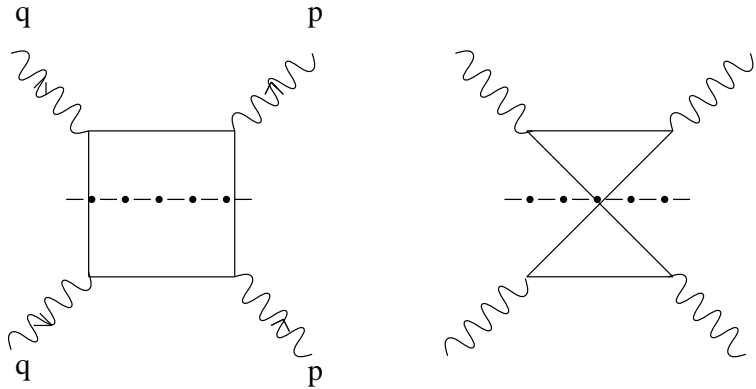,width=10cm}
\caption{Fermion loop contribution to the imaginary part of the
$\gamma^*\gamma$ forward scattering amplitude.}
\end{center}
\end{figure}  

The lowest order fermion loop does not contribute to the anomalous dimension.
Notice that the resulting amplitude (\ref{2.1}), (\ref{2.2}) is even in
exchanging $s$ and $u$ channels. We see that the positive signature amplitude
 is relevant for the photon spin structure function.

It is known that contributions to the anomalous dimensions ($Q^2$ evolution)
arise from the exchange of two gluons and these contributions are
proportional to $x$, i.e. they dominate (\ref{2.2}) at small $x$
\cite{A-M}, \cite{Schmidt}.

We calculate the lowest order contribution with two gluon exchange in the
asymptotics of small $x$ (Fig. 2).

\begin{figure}[htb]
\label{ffig2}
\begin{center}  
\epsfig{file=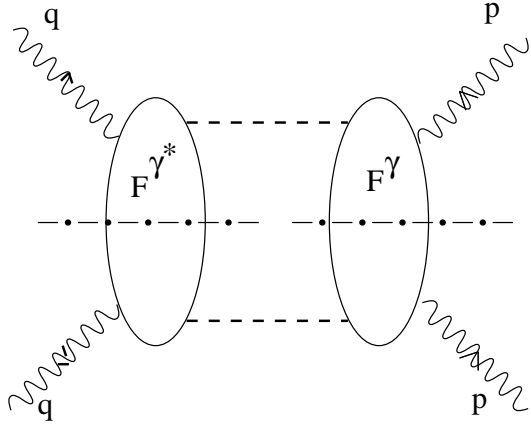,width=7cm}
\caption{Two-gluon exchange contribution to $\gamma^*\gamma$ forward
scattering.}
\end{center}
\end{figure}  

As building blocks we have the
imaginary parts of the fermion loop photon-gluon scattering amplitude,
on one side with the virtual photon $q$ and on the other with the real photon
$p$.
Both can be read off from \cite{Zima}, where the same projector (\ref{2.1})
applies ($k_\mu = -\al q'_\mu + \be p_\mu + \ka_{\perp \mu},\;\;
q'_\mu~=~q_\mu~-~\frac{q^2}{2pq}p_\mu$)
\beqar{2.3}
&&F^{\ga^*}(\be\,s,\ka_\perp;Q^2) = 8\,e^2g^2\,\frac{\ka_\perp^2 Q^2}{(s\be)^2}
\ln \frac{s\be}{\sqrt{\ka_\perp^2 Q^2}}\;, \;\;\;Q^2 \gg |\ka_\perp^2|\;,
 \nonumber \\
&&F^{\ga}(\al\,s,\ka_\perp;Q_0^2) = -2\,e^2g^2 \frac{(\ka_\perp^2)^2}{(s\al)^2}
\;, \;\;\;Q_0^2 \ll |\ka_\perp^2|\;.
\eeqar
$F^{\ga}$ for $Q_0^2 \ll |\ka_\perp^2|$ takes into account the direct photon 
contribution. A resolved (non-perturbative) contribution is to be added.
The results are approximate according to the Regge kinematics appropriate
at small $x$:
\beqar{2.4}
&& s\al \be \ll \ka_\perp^2\;, \;\;\;\;s \gg \ka_\perp^2\;, \nonumber \\
&& s\al \gg \ka_\perp^2 \;, \;\;\;\;s\be \gg \ka_\perp^2 + Q^2\;\;.
\eeqar
We obtain $(x=\frac{Q^2}{s})$
\beqar{2.5}
&&\int \frac{d^4\,k}{(k^2 + i\epsilon)^2}F^{\ga^*}(\be s,\ka_\perp;Q^2)
F^{\ga}(\al s,\ka_\perp;Q_0^2) \nonumber \\
 &&= -e^4g^4N^2\left[2\pi x \ln^2 x + 8\pi x \left(\frac{\pi^2}{12} -1  
\right) + {\cal O}(x^2)\right]\;\;.
 \eeqar
The resulting contribution to the structure function behaves indeed like
$x$ with logarithmic corrections.

The $Q^2$ evolution arises from the interaction of the exchanged gluons
with strong ordering of the transverse momenta \cite{DGLAP} (see Figs. 3a,b)
\beqar{2.6}
&&  F_3^\ga =  F^{(0)}_{3}  \otimes  F_3^g \;\;, \nonumber \\
&& \frac{d}{d\ln Q^2}\;F_3^g(x,Q^2) = \frac{g^2(Q^2)}{8\pi^2}
\;\int\limits^{1}_{x}\,\frac{dz}{z}\,P(\frac{z}{x})\,F_3^g(z,Q^2)\;\;.
\eeqar
The photon structure function is obtained from the gluon transversity
$F_3^g$ by convolution with the coefficient function proportional to 
$F^{(0)}_{3}$ (\ref{2.2}).

\begin{figure}[htb]
\label{ffig3}
\begin{center}  
\epsfig{file=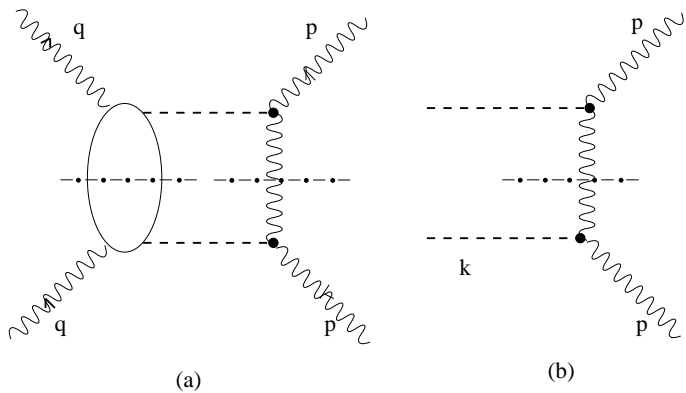,width=9cm}
\caption{One-loop contribution to the $Q^2$ evolution.}
\end{center}
\end{figure}

The 
initial-value $x$-distribution at $Q_0^2$ for eq.(\ref{2.6})
 is the sum of the perturbative contribution
of $F^{\ga}$ (\ref{2.3}) and a non-perturbative resolved contribution.
The evolution kernel $P(z)$ is readily obtained, e.g. by calculating the
graph in Fig. 3b in the axial gauge, $A^\mu q'_\mu =0$,
\beq{2.7}
    P(z) = \frac{2Nz}{1-z} \;.
\eeq
The anomalous dimensions corresponding to this
  result has been presented first in
\cite{Bukhvostov} without relation to $F_3^\ga$. Later this splitting function
was derived in  \cite{A-M} and in
\cite{Schmidt}. In (\ref{2.7}) we do not write the contributions
$\sim \de(z-1)$ which are irrelevant for the small-$x$ asymptotics.

We shall see that this leading $\ln Q^2$ evolution leads to a
small-$x$ behaviour of $F_3^\ga$ proportional to $x$ up to logarithmic
corrections. Correspondingly, the one-loop 
anomalous dimension has a pole at $j=-1$.
The logarithmic correction to the small-$x$ asymptotics are not
completely accounted for by (\ref{2.7}) as we shall discuss below.

\section{Double log contributions at small $x$}
\setcounter{equation}{0}

\subsection{The gluon ladder graphs}

We consider the contribution of a $s$-channel intermediate state of gluons
in multi-Regge kinematics to the interaction of the exchanged gluons (see
Fig. 4a).

\begin{figure}[htb]
\label{ffig4}
\begin{center}  
\epsfig{file=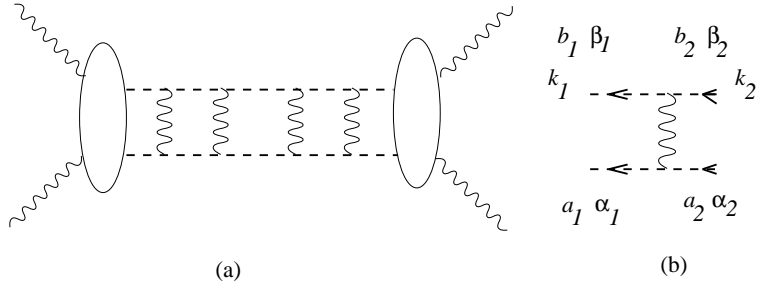,width=10cm}
\caption{The gluon ladder contribution.}
\end{center}
\end{figure}

From the DGLAP equation (\ref{2.6}) we see that there is a double log
contribution
\[ \sim g^2 (\ln \frac{1}{x}) \ln Q^2 \]
from each loop arising from the configuration of strongly ordered transverse
momenta. We show that there are further double logarithms
resulting in a contribution $\sim g^2\ln^2 \frac{1}{x}$ from each loop
which determine the small-$x$ behaviour.

As the first step we analyze the one-rung contribution (Fig. 4b)
in the appropriate projection (\ref{2.1})
\beqar{3.1}
&& \Ga^{\al_2 \al_1 \si}(k_2,k_1)f^{a_2 a_1 c} \;g^\perp_{\si \si'}
\Ga^{\si' \be_1 \be_2}(k_1,k_2)f^{c b_1 b_2} \nonumber \\
&& \frac{\de_{a_1 b_1}\de_{a_2 b_2}}{N^2 - 1}
(P_{(+-;-+)} +
P_{(-+;+-)})_{\al_1 \be_1,\al_2 \be_2} \nonumber \\
&& = 4N(\ka_{1\perp}^2 + \ka_{2\perp}^2)\;\;,
\eeqar
where $\Ga$ is the triple-gluon vertex.
Note that only the transverse gluon polarizations contribute to the
$s$-channel intermediate states.

In order to have a double log contribution from each loop we have to
obtain a logarithmic contribution from the transverse momentum
$\ka_{i\perp}$ integral. For this a factor of $\ka_{i\perp}^2$ 
from the numerator
as is provided by the result (\ref{3.1}) is essential. The coefficient $4N$
in (\ref{3.1}) determines the size of the double log contribution.

The ladder graphs shown in Fig. 4a are summed by an integral equation
the kernel of which is obtained in an obvious way using (\ref{3.1}).
The peculiarity are the limits of integration in the longitudinal momentum
fraction $\be$ and the transverse momentum $\ka$, which show whether
there are double logs beyond the strong ordering region in $\kappa$:
\beq{3.2}
x\ll \dots \be_i \ll \be_{i+1} \dots \ll 1\;,\;\;\;\;
\frac{Q^2}{x}\gg \dots \frac{|\ka_i^2|}{\be_i} \gg
\frac{|\ka_{i+1}^2|}{\be_{i+1}} \dots \gg \mu^2\;\;.
\eeq
Here and in the following we denote the transverse vectors
by means of  the complex number
 ($\ka = \ka_{\perp}^1 + i\ka_{\perp}^2$).
Indeed we find that there is strong ordering in $\be$'s and in
$\frac{|\ka^2|}{\be}$, which leads to a double log region larger
than the one in the DGLAP equation.

The discussion above relies on the simple ladder graphs assuming that
there are no further double log contributions. This is indeed the true
for the negative signature channel, but not obvious.
However we finally have to calculate the positive signature partial wave,
as we have pointed out in Sec. 2. We apply the method of Ref.~\cite{KL}.

\subsection{The soft $t$-channel intermediate state}

The leading $\ln s$ approximation in the considered channel results
in a sum of ladder graphs summed by an BFKL-like equation. The ladders
are built of effective vertices and reggeized gluons and each of them
is a gauge invariant sum (in the considered approximation) of Feynman
graphs. The leading $\ln s$ contributions in the vicinity of $j=-1$ and
for negative signature, i.e. the $\ln s$ corrections to the power
$s^{-1}$ contribution to the amplitude, include in particular the double
log contributions in question. Consider the transverse momentum
$\ka_i$ integral in one of the ladder loop. The region where
$\ka_i$ is smaller than all other transverse momenta in the loops is
to be investigated in particular. The question about the double log
beyond the DGLAP equation reduces to the question of whether the transverse
momentum integral is logarithmic in this region of smallest $\ka$.

\begin{figure}[htb]
\label{ffig5}
\begin{center}  
\epsfig{file=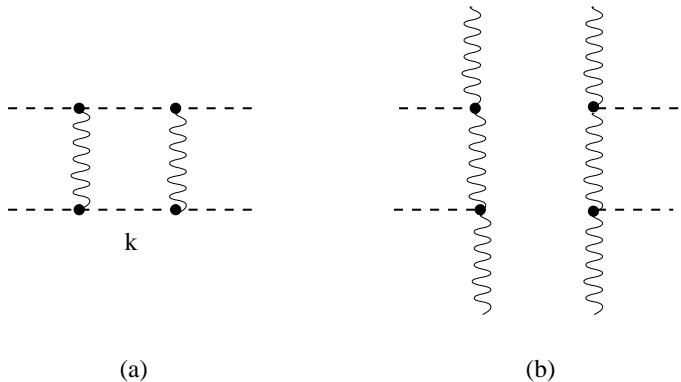,width=9cm}
\caption{Gluon ladder element with the smallest transverse momentum (a) and
the related gluon scattering graphs (b).}
\end{center}
\end{figure}

The ladder loop at $\ka \to 0$, Fig. 5a, is related to the product of
graphs with scattering gluons Fig.~5b. In this way the exchanged ($t$-channel)
reggeized gluons are related to scattered ($s$-channel) gluons.
Due to the particular  
projection (\ref{2.1}) the gluon scattering is accompanied
by helicity flip.

The scattering graph Fig. 5b stands for gauge invariant sum of graphs
involving effective scattering vertices. In the tree approximation there
are no effective scattering vertices changing helicity with leading (near
$j=1$) or subleading (near $j=0$) gluonic reggeons \cite{gr}. 
A helicity-flip
effective vertex appears only at the next-to-subleading ($j=-1$) level.
We conclude that the contribution to the considered channel (\ref{2.1})
arises from the exchange of one leading ($j \sim 1$) and one
next-to-subleading ($j \sim -1$) gluonic reggeon.

Now we consider again the relation of the scattering graphs Fig. 5b
to the ladder loop Fig. 5a ($t$-channel unitarity).
In order to get Fig. 5a from Fig. 5b one has to turn the scattering
gluons into exchanged reggeons and add the propagators.
The trajectory functions $\al(\ka)$ can be disregarded in the double log
approximation.

The relation of a leading gluonic reggeon to a scattering gluon in
the small $\ka$ region has been analysed in \cite{LKSsym}. It is
important that there arises a factor of $\ka$ at each vertex in turning
the scattering gluon into the leading gluonic reggeon.
The subleading gluonic reggeon is obtained from the scattering gluon
without such additional factor. As a result we have a factor $|\ka|^2$
and  together with the two propagators we obtain a logarithmic transverse momentum
integral.

We compare the situation to the other channels:
In the case of two leading gluonic reggeons we would have instead a
factor $|\ka|^4$ and in the case of two subleading reggeons no additional
factors of $\ka$. In both cases no double log contribution from the soft
$\ka$ region appears.

The scattering graphs Fig. 5b are directly related to the parton
splitting kernel $P(z)$ in the DGLAP equation (\ref{2.6}). Each of the
scattering graphs can be compared with the graph Fig. 3b, from which
we have obtained $P(z)$ using axial gauge $q'A=0$. Indeed, the gauge
invariant effective scattering vertices coincide with the bare vertices in
the corresponding axial gauge.

The double
log contribution of the BFKL ladders can be summed by the equation given
graphically in Fig. 6.

\begin{figure}[htb]
\label{ffig6}
\begin{center}  
\epsfig{file=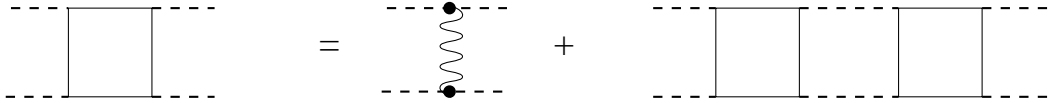,width=14cm}
\caption{Equation summing the double log contribution from soft $t$-channel
intermediate states.}
\end{center}
\end{figure}  

Introducing the partial waves we can write it in
the following form ($\om =j+1$)
\beq{3.3}
    f^{(-)}(\om) = \frac{g^2a_0}{\om} +
\frac{1}{8\pi^2}\frac{1}{\om}\,(f^{(-)}(\om))^2\;\;.
\eeq
The Born term corresponds in fact to one of the scattering graphs in Fig. 5b;
$\frac{a_0}{j+1}$ is the leading contribution at $j\to -1$ of the moment
transformation of $P(x)$
\beq{3.4}
\frac{a_0}{j+1} = \int\limits_0^1 d\,x\; x^{j-1}\;P(x)\;|_{j\to -1}
\eeq
and we have $a_0=2N$. \eq{3.3} is readily solved
\beq{3.5}
f^{(-)}(\om) = 4\pi^2 \om \left( 1- \sqrt{1-
\frac{4\al_S\,N}{\pi \om^2}} \right)\;\;.
\eeq
However this is not the final result for the double log contributions
to the parton spin-flip amplitude (\ref{2.1}) because we have
calculated so far only the negative signature contribution.

\subsection{Soft bremsstrahlung contributions}

The leading $\ln s$ effective ladder equation of BFKL type is obtained
under the essential assumption of colour singlet state in $t$-channel.
Only in this case the infrared divergencies cancel.
To analyze amplitudes and partial waves corresponding to non-singlet channels
we restrict the integrations over the transverse momentum $\ka$ by the
condition $|\ka|^2 > \mu^2$. In non-singlet channels there are more
double log contributions besides of those contained in the ladder.
Loops with a single gluon being soft, i.e. carrying the smallest
$\ka$, give rise to double logs. This is different from the two-gluon
$t$-channel intermediate state discussed above. The contribution of the
soft gluon loop is easily calculated relying on Gribov's bremsstrahlung
theorem: The gluon  with the smallest transverse momentum is effectively
emitted and absorbed from the external lines. With the bremsstrahlung
contribution the equation in Fig. 6 generalizes to the one shown in Fig.7.

\begin{figure}[htb]
\label{ffig7}
\begin{center}  
\epsfig{file=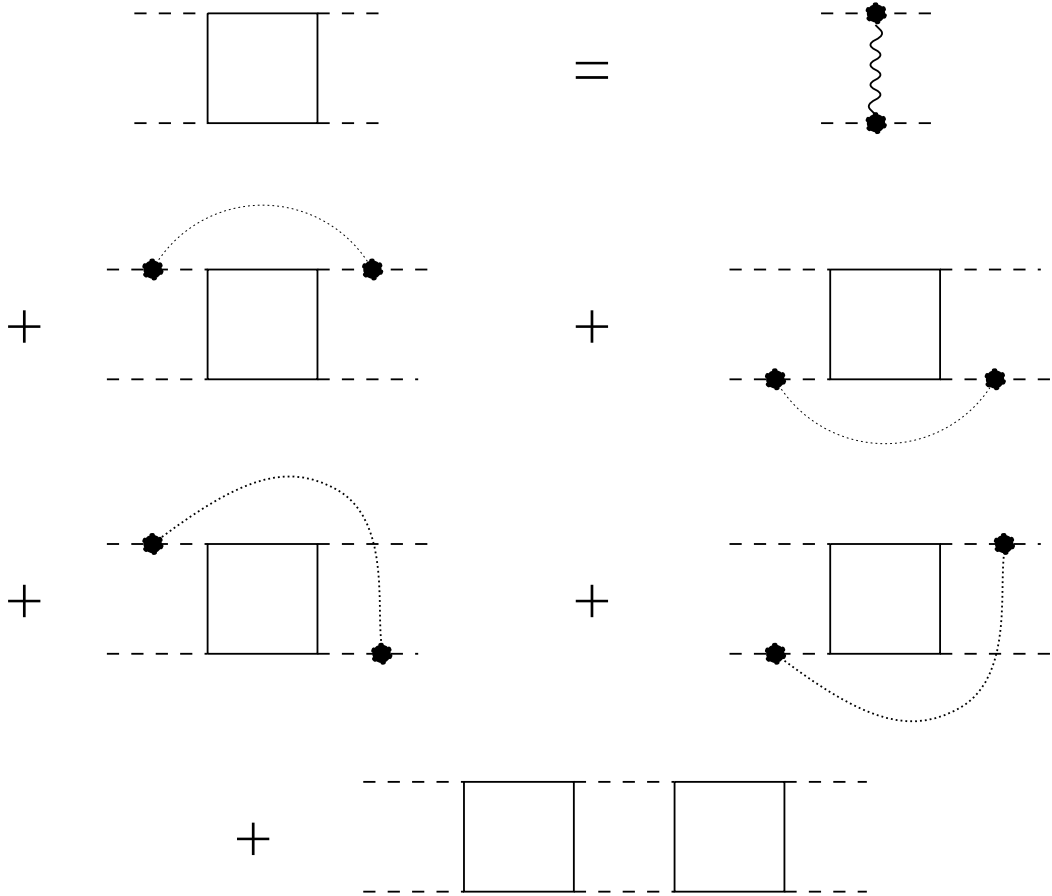,width=14cm}
\caption{Equation summing the double log contributions both from 
soft $t$-channel intermediate states and from soft bremsstrahlung.}
\end{center}
\end{figure}

The soft bremsstrahlung contributions change the gauge group quantum 
numbers in the $t$-channel.
Now instead of a single partial wave $f(\om)$ we have to consider 
a column vector
involving the colour singlet partial wave discussed so far together
with the colour octet and further colour channnels
\beq{3.6}
    \hat{f}= \left(
    \begin{array}{c}
    f \\ f_8 \\ \cdot \\ \cdot \\ \cdot
    \end{array}
    \right)\;\;.
\eeq
We shall concentrate on these first two entries, disregarding the remaining
representations which appear in the symmetric part of the tensor product 
of two adjoint (gluon) representations..
Let us notice  that the octet contribution in question is the one
arising by antisymmetrization, i.e. by projection with $f^{abc}$.

In colour space the interaction induced 
by gluons attached to the external lines is expressed by multiplying the column
of partial waves by a matrix. This matrix depends on whether the pair of
those external lines belongs to the $s$- or $t$- channels. The upper limits of the
double logarithmic 
transverse momentum integrals are respectively $s$, $u$ or $t$.
Therefore the $t$-channel type soft gluons are not important in our case.
Nevertheless it is reasonable to introduce all three matrices.
Obviously $\hat{M}_t$ is diagonal
in the representation with definite quantum numbers in the $t$-channel.
 Restricting to the two relevant
colour channels we have
\beq{M_t}
\hat{M}_t = \left( \begin{array}{cc}
N & 0 \\
0 & \frac{N}{2}
\end{array}
  \right) \;\;\;.
\eeq
The sum of the three types of colour interactions taken with the same weight
(the interaction in momentum space leads to different weights, however)
acts as the identity up to the factor $N$
\beq{sum_M}
\hat{M}_s + \hat{M}_u + \hat{M}_t = N\,\hat{I}
\eeq 
Working out the relations of $SU(N)$ representations leads to
\beq{M_s}
\hat{M}_s = \left( \begin{array}{cc}
0  & N \\
\frac{N}{N^2 - 1} & \frac{N}{4}
\end{array}
  \right) \;\;\;.
\eeq
The complete matrices for all colour representations appearing in the 
two-gluon exchange are given in \cite{FK}.

The soft bremsstrahlung contribution is readily calculated in momentum space.
The sum of 4 diagrams, two of $s$-channel and two of $u$-channel type, 
results in
\beq{3.7}
\frac{g^2}{4\pi^2} \int\limits_{\mu^2}^s\;\frac{d\,|\ka|^2}{|\ka|^2}\,
\left[ \hat{M}_s \ln (\frac{-s }{|\ka|^2}) +
\hat{M}_u \ln (\frac{s }{|\ka|^2}) \right] \hat{A}^{(\sigma)}(s,|\ka|^2)\;\;.  
\eeq
 The square bracket in (\ref{3.7}) involves an odd contribution in $s$, which
leads to a change  in signature $\sigma$.
Here $\hat{A}^{(\sigma)}(s,|\ka|^2)$ is the column vector of (almost
on-shell) gluon 
forward scattering amplitude where the double log corrections are
included with $\ka$ as the lower cut-off in all transverse momentum
integrals
\beq{A}
\hat{A}^{(\sigma)}(s,|\ka|^2) = \int\limits^{i\infty}_{-i\infty}
 \frac{d\,\om}{2\pi i} 
\hat{f}^{(\sigma)}(\om) \left(\frac{s}{|\ka|^2}\right)^\om \; 
{\zeta}^{(\sigma)}(\om)\;\;. 
\eeq
Taking into account the signature factor approximated for small $\om=j+1$
\beq{zeta}
{\zeta}^{(-)}(\om)\approx 1\;\;,\;\;\;\;\; 
{\zeta}^{(+)}(\om)\approx -\frac{1}{2}\,i\pi\om\;,
\eeq
we have in terms of partial waves
\beqar{f}
&&\hat{f}^{(-)}(\om)\Bigg|_{brems} = \frac{g^2}{4\pi^2}\left( \hat{M}_s + \hat{M}_u
\right) \frac{1}{\om}\frac{d}{d\om} \hat{f}^{(-)}(\om) \nonumber \\
&&\hat{f}^{(+)}(\om)\Bigg|_{brems} =  \\
&&\frac{g^2}{4\pi^2}\left( \hat{M}_s + \hat{M}_u
\right) \frac{1}{\om^2}\frac{d}{d\om}(\om \hat{f}^{(+)}(\om)) -
\frac{g^2}{4\pi^2}\frac{1}{\om^2}\left( \hat{M}_s - \hat{M}_u \right)
\hat{f}^{(-)}(\om)\;\;.\nonumber
\eeqar
The signature changing contribution is relevant in our approximation 
(small $\om$) only if multiplied with the negative signature amplitude 
and 
therefore leads to an additional inhomogeneous term in the positive signature
equation.

We write the resulting partial wave equations in matrix form
\beqar{3.9}
&&\hat{f}^{(-)}(\om) = a\hat{M}_t\frac{g^2}{\om} + \frac{g^2}{4\pi^2}
\left( \hat{M}_s 
+ \hat{M}_u \right)\frac{1}{\om}\frac{d}{d\om}\hat{f}^{(-)}(\om)
+ \frac{1}{8\pi^2 }\frac{1}{\om}(\hat{f}^{(-)\,2}(\om)) \nonumber \\
&&\hat{f}^{(+)}(\om) =
a\hat{M}_t \frac{g^2}{\om} + \frac{g^2}{4\pi^2} 
\left( \hat{M}_s + \hat{M}_u \right)\frac{1}{\om^2}
\frac{d}{d\om}(\om \hat{f}^{(+)}(\om)) \nonumber \\
&&- \frac{g^2}{4\pi^2}\left( \hat{M}_s - \hat{M}_u \right)\frac{1}{\om^2}
\hat{f}^{(-)}(\om)
+ \frac{1}{8\pi^2 \om}(\hat{f}^{(+)\,2}(\om)) \;.
\eeqar
$(\hat{f}^{(\pm)\,2})$ denotes the column vector with the squares of the partial waves
of the corresponding colour channel. Comparing with $a_0$ in (\ref{3.3})
for the colour singlet channel we identify $a=2$.

This is the part of the residue $a_0$ of the one-loop anomalous
dimensions at $j=-1$ not related to colour factors but merely to the
helicity state. Comparing to the gluonic contribution to the double
log small-$x$ asymptotics of the helicity asymmetry $g_1$ \cite{g1} 
and to the flavour non-singlet (quark-antiquark) contributions \cite{h1}
we see that there we have $a=4$ and 1, respectively.

Since the matrix element $(\hat{M}_s + \hat{M}_u)_{11}$ vanishes
the equation for the colour singlet component for both signatures is just
an algebraic equation of second order. The negative signature solution has 
been discussed above. For the positive signature we obtain
\beq{f+}
f^{(+)}_0 = 4\pi^2 \om \left( 1 - \sqrt{1 - \frac{2\al_S\,N}{\pi \om^2}
\left(2 - \frac{1}{2\pi^2 \om}f^{(-)}_8(\om)   \right)} \right)\;\;.
\eeq
The solution is expressed in terms of the negative signature octet partial
wave. The corresponding equation is differential of Ricatti type.
The solution can be expressed in terms of the logarithmic derivative
of the parabolic cylinder function ${\cal D}_p(z)$ \cite{KL}
\beqar{f-8}
&&f^{(-)}_8 = 4\pi \al_S N \frac{d}{d\,\om} \ln \left(
  \exp(\frac{\om^2}{4\bar{\om}^2}){\cal D}_p(\frac{\om}{\bar{\om}})
\right)\nonumber \\
&&\bar{\om}^2 = \frac{\al_S\,N}{2\pi}\;,\;\;\;p=2\;\;.
\eeqar
The small-$x$
dependence of $F_3^\ga(x, Q^2)$ is obtained from the partial wave
$f^{(+)}_0$
by inverse Mellin transform
\beq{inverse}
F_3^\ga(x, Q^2) \sim \int\limits^{i\infty}_{-i\infty} \frac{d\,\om}{2\pi i}\,
f^{(+)}_0(\om) \zeta^{(+)}(\om)\; x^{-\om} \;\;.
\eeq
In the relation we did not include explicitely the convolution of the
reggeon Green 
function $f^{(+)}_0$ with the appropriate impact factors, which is not
important neither for the leading small-$x$ behaviour nor for the
leading term of the anomalous dimensions. The small-$x$ asymptotics
is determined by the right-most singularity $\om^{(+)}_0$ of $f^{(+)}_0(\om)$
\beq{smallx}
F_3^\ga(x, Q^2)\Bigg|_{double-log} \sim x^{1-\om^{(+)}_0}\;\;.
\eeq
$\om^{(+)}_0$ is close to $\om^{(-)}_0$, the branch point of $f^{(-)}_0(\om)$
\beq{omega-}
\om^{(-)2}_0 = \frac{4\al_S\,N}{\pi} \;\;.
\eeq
Assuming $\al_S=0.2$, $N=3$ we have $\om_0^{(-)}\approx 0.87$.
As a rough estimate of the signature changing contribution
we replace $f_8^{(-)}$ by its Born term. Unlike the case of two fermion
exchange the latter is not suppressed in the large $N$ 
approximation, therefore this estimate is worse in the present case.
We obtain $\om_0^{(+)} \approx 0.61$.

In the double-log approximation also the resummed anomalous
dimensions $\nu^{(+)}(j)$ near $j=-1$ are obtained in terms of the partial wave
$f^{(+)}_0(\om)$, $\om = j+1$,
\beq{nu}
\nu^{(+)}(-1+\om) = \frac{1}{8\pi^2}\;f^{(+)}_0(\om) \;\;.
\eeq

\section{The perturbative pomeron contribution}
\setcounter{equation}{0}

We calculate in the leading $\ln s$ approximation the contribution to the
virtual photon spin-flip amplitude (\ref{2.1}) with the asymptotics
$s^{1+\om_2}$ leading to a contribution to the structure function behaving
like $x^{-1 -\om_2}$. 
There are well known results for the real and virtual photon impact factors
with the exchange of two leading gluonic reggeons,
describing the coupling of the scattering photon to the BFKL pomeron \cite{BFKL}
via a quark loop \cite{impactf}, \cite{BL-borel}, \cite{L-conf}. 
These impact factors involve a helicity-flip 
contribution at vanishing momentum transfer.

For a virtual photon the  helicity-flip
($\Delta \la=\pm 2$) contribution 
to the impact factor
reads 
\beqar{impactf}
&&\Phi^{(+,-)}_A(\ka, Q^2) = \frac{\al_{e.m.}\al_S}{\sqrt{2\;\pi}}\cdot 4
\sum\limits_{q}\int\limits^1_0 dx \int\limits^1_0 dy 
\;\frac{x(1-x)y(1-y)\,\ka^{*2}}{|\ka|^2x(1-x) + Q^2y(1-y) + m_q^2} 
\nonumber \\
&& \Phi^{(-,+)}(\ka, Q^2) = \left(\Phi^{(+,-)}(\ka, Q^2)\right)^* \;, 
\;\;\;\;\;Q^2\gg m_\rho^2 \;\;  .
\eeqar
For quasi-real photons $(Q^2 \leq m_\rho^2)$ this expression is applicable
only for the heavy flavour contribution. Non-perturbative contributions are
essential and can be roughly described by applying the Borel transform 
with respect to $Q^2$ to (\ref{impactf}) and identifying the variable
conjugated to $Q^2$ with the $\rho$-meson mass $m_\rho$ \cite{BL-borel}, 
\beqar{impactf-borel}
&&\Phi^{(+,-)}_B(\ka, Q^2) 
= \frac{\al_{e.m.}\al_S}{\sqrt{2\;\pi}}\cdot 4\cdot N_q\cdot 
\int\limits^1_0 dx \int\limits^1_0 dy \frac{x(1-x)\ka^{*2}}{m_\rho^2}
\exp {\left(-\frac{|\ka|^2\,x(1-x)}{m_\rho^2\,y(1-y)} \right)} \nonumber \\
 && + H.Q.\;\;,\;\;\;\;Q^2 \leq m_\rho^2 \;\;.
\eeqar
Here $N_q$ is the number of light quarks ($N_q=3$) and $H.Q.$
denotes the expression (\ref{impactf}) with the sum now restricted to 
the heavy quark flavours.

In the Regge asymptotics the partial wave of the photon spin-flip forward
scattering amplitude can be calculated from the impact representation
\beq{im-repr}
f_{+,-;-,+}(\om,Q^2) = \int\, 
\frac{d^2\,\ka_1\;\;d^2\,\ka_2}{|\ka_1|^4\;|\ka_2|^4}
\Phi^{(+,-)}_A(\ka_1, Q^2)\;F(\om;\ka_1,\ka_2) \;\Phi^{(-,+)}_B(\ka_2, Q_0^2)\;\;.
\eeq
In our case
the reggeon Green function $F(\om;\ka_1,\ka_2)$ 
is  obtained from the well
known solution of the BFKL equation \cite{L-conf}, \cite{BFKL}. In the sum over the 
conformal spin $n$ only the term $n=2$ contributes
\beqar{n2}
&&F(\om;\ka_1,\ka_2) = \frac{1}{2\pi^2}\int\;
\frac{d\nu\;|\ka_1^2|^{-\frac{1}{2}+i\nu} \; |\ka_2^2|^{-\frac{1}{2}-i\nu} 
\ka_1^2\; \ka_2^{*2}}{\om - \frac{g^2\,N}{8\pi^2}\Omega(2,\nu)}\;\;, \nonumber \\
&&\Omega(n,\nu) = 4\psi(1) - \psi(\frac{1}{2}+i\nu + \frac{n}{2})
-\psi(\frac{1}{2}-i\nu + \frac{n}{2}) \nonumber \\
&&- \psi(\frac{1}{2}+i\nu - \frac{n}{2})
-\psi(\frac{1}{2}-i\nu - \frac{n}{2}) \;\;.
\eeqar
The spin-flip partial wave has a cut with the right branch point at angular
momentum $j=1+\om_2$,
\beq{omega2}
\om_2 = \frac{g^2\,N}{8\pi^2}\Omega(2,\nu) = 
\frac{g^2\,N}{\pi^2}(\ln 2 - 1)\;\;.
\eeq
Choosing $\al_S=\frac{g^2}{4\pi}=0.2$ 
as in the usual estimate for the BFKL pomeron
intercept and $N=3$ we have $\om_2 = -0.23$.

Investigating the $Q^2$ behaviour of (\ref{n2}) we find that this 
perturbative pomeron contribution does not influence the anomalous
dimension (in the vicinity of moment number $j=1$) at the leading
$\ln Q^2$ level. Its influence on the anomalous dimension arises at the
next-to-leading level (as a pole term $\sim \frac{1}{j-1}$) from the 
iteration of both considered contributions. 

\section{Discussion}
\setcounter{equation}{0}

There are two contributions to the small-$x$ asymptotics of the polarized
photon structure function $F_3^\ga(x,Q^2)$
\beq{d1}
F_3^\ga(x,Q^2) = F_3^\ga(x,Q^2)\Bigg|_{double-log} +  
F_3^\ga(x,Q^2)\Bigg|_{BFKL}\;\;.
\eeq
The first is closely related to the DGLAP evolution. The small-$x$
asymptotics is obtained in the double logarithmic approximation 
extending the double-log contribution to the DGLAP equation
beyond the region of strong ordering  in the transverse momenta.
The $t$-channel partial wave describing this contribution and the 
related anomalous dimension near angular momentum $j=-1$
in all orders of perturbation theory have been calculated.
The small-$x$ behaviour of this contribution is found to be
$x^{1-\om_0^{(+)}}$, where the displacement $\om_0^{(+)}$ is of 
order $\sqrt{\al_S}$ and is estimated to be $\om_0^{(+)} \approx 0.6$.

The second contribution is not directly related to the DGLAP
evolution. Its trace appears in the anomalous dimensions as pole 
terms at $j=+1$ starting from the two-loop approximation.
This contribution arises from the conformal spin $n=2$ term of
the BFKL pomeron solution \cite{L-conf}, \cite{BFKL}. It is a term which is not 
essential in the usual phenomenological applications of the 
BFKL pomeron. Therefore the experimental study of $F_3^\ga(x,Q^2)$
would allow to test the detailed structure of the BFKL pomeron.
This contribution dominates at small $x$ and behaves like
$x^{-1-\om_2 }$, where the displacement $\om_2$ is of order ${\al_S}$
and is estimated to be $\om_2 \approx - 0.23$.

Both contributions arise from the exchange in $t$-channel of two
(reggeized) gluons interacting by $s$-channel gluons. $s$-channel
helicity $\si=2$ is transferred in both cases.

In the first case one of the reggeized gluons is the leading one 
carrying $\si=0$ and the other a twice-subleading one carrying
$\si=2$. The latter couples to gluons scattering with 
helicity flip, whereas the first does not feel the helicity
of scattering partons. The resulting Regge singularity is a branch cut at
$j=-1+\om^{(+)}_0$.
In the second case both reggeized gluons are the leading ones ($\si=0$),
which are, however, in a state with the longitudinal projection of the orbital angular momentum
equal to 2. The resulting Regge singularity is a branch cut at
$j=1+\om_2$.

We notice that also in the small-$x$ asymptotics of the structure function
$h_1(x,Q^2)$ measuring the quark transversity we have encountered
two contributions in analogy to (\ref{d1}). There we have two reggeized
quarks in the $t$-channel which carry $s$-channel helicity
$\si=1$. The first contribution ($\sim x$ plus corrections)
related to the DGLAP evolution, arises from the exchange of one leading 
and one subleading quark reggeon. The second contribution (constant in $x$)
arises from two leading quarks with parallel helicities.

We have given the couplings (impact factors) of these exchanges to the
scattering photons explicitely. We did not perform here a numerical analysis
of the expressions obtained for the amplitude related to 
$F_3^\ga(x,Q^2)$. It is clear, however, that the couplings of the angular 
momentum 2 state are weaker. 
Therefore we expect that the first contribution
($\sim x^{1-\om^{(+)}_0}$) will  dominate at not too small values of
$x$ and will be overcome by the second ($\sim x^{-1-\om_2}$) only at very 
small $x$.

\vspace*{1cm}
{\Large \bf  Acknowledgments}

\vspace*{.8cm}

L.Sz. would like to acknowledge the warm hospitality extended to him at
University of Leipzig where this work was completed.


\begin{thebibliography}{99}

\bibitem{Witten} E.~Witten, Nucl. Phys. B120(1977)189


\bibitem{photon}
S. S\"oldner-Rembold, Talk at XVIII Intern. Symposium on Lepton Photon
Interactions, Hamburg, Germany, July 28--August 1, 1997, hep-exp/9711005; \\
R. Nisius, Talk at the Intern. EPS Conference on High Energy Physics,
19-26 August 1997, Jerusalem, Israel, hep-exp/9712012; \\
M. Krawczyk, M. Staszel and A. Zembrzuski, "Survey of recent data on 
photon structure functions and resolved photon processes", hep-ph/9806291


\bibitem{f-loop} R. Karplus and M. Neuman, Phys. Rev. 80(1950)350; \\
V.N. Baier, V.S. Fadin and V.A. Khose, Sov. Phys. JETP 23(1966)104; \\
L. Chernyak and V.G. Serbo, Nucl. Phys. B71(1974)395\\
M.A. Ahmed and G.G. Ross, Phys. Lett. 59B(1975)369 



\bibitem{HJM} P.~Hoodbhoy, R.J.~Jaffe and A.~Manohar, Nucl. Phys. 
B312(1989)571; \\
A.~Manohar, Phys. Lett. 219B(1989)357



\bibitem{J-M89} R.L.~Jaffe and A.~Manohar, Phys. Lett. B223(1989)218

\bibitem{transversity}
J.~Ralston and D.E.~Soffer, Nucl. Phys. B152(1979)109; \\
R.L.~Jaffe and X.~Ji, Phys. Rev. Lett. 67(1991)553; 
Nucl.~Phys.~B375(1992)527

\bibitem{h1}
R.~Kirschner, L.~Mankiewicz, A.~Sch\"afer and L.~Szymanowski,
Zeit.~f.~Physik C74(1997)501 

\bibitem{h1Q}
S.~Kumano and M.~Miyama, Phys. Rev. D56(1997)2504;\\
A.~Hayashi, Y.~Kanazawa and Y.~Koike, Phys. Rev. D56(1997)7350;\\
W.~Vogelsang, Phys. Rev. D57(1998)1886

\bibitem{HERA} Proceedings of the Workshop "Deep-inelastic scattering off
polarized targets: Theory meets experiment", eds. J.~Bl\"umlein,
A.~de~Roeck, T.~Gehrmann and W.-D.~Nowak, DESY-97-200

\bibitem{eff-action}
L.N.~Lipatov, Nucl. Phys. B365(1991)614; \\
R.~Kirschner, L.N.~Lipatov and L.~Szymanowski, Nucl. Phys. B425(1994)579

\bibitem{LKSsym} R. Kirschner, L.N. Lipatov and L. Szymanowski,
Phys. Rev. D51(1995)838



\bibitem{gr} R. Kirschner and L. Szymanowski, Phys. Lett. B419(1998)348;
and hep-ph/9712456, in print in Phys. Rev. D58 (1998)

\bibitem{Asimov}
Ya.I. Asimov, ZETF 43(1962)2321; \\
S.~Mandelstam, Nuovo Cim. 30(1963)1127;\\
V.N.~Gribov, I.Ja. Pomeranchuk and K.A. Ter-Martirosyan, Yad.~Fiz.~2(1965)361


\bibitem{Zima} V.G. Zima, Sov. J. Nucl. Phys. 16(1973)580

\bibitem{A-M} X.~Artru and M.~Mekhfi, Zeit. f. Physik C45(1990)669

\bibitem{Schmidt} E.~Sather and C.~Schmidt, Phys. Rev. D42(1990)1424


\bibitem{Bukhvostov} A.P. Bukhvostov, G.V. Frolov, E.A. Kuraev and
L.N. Lipatov, Nucl. Phys. B258(1985)601




\bibitem{DGLAP}
V.G. Gribov and L.N. Lipatov, Sov. J. Nucl. Phys. 15(1972)438 \\
L.N. Lipatov,  Yad. Fiz. 20(1974)532     \\
G.~Altarelli and G.~Parisi, Nucl. Phys. B126(1977)298 \\
Yu.L. Dokshitzer, ZhETF 71(1977)1216




\bibitem{KL} R. Kirschner and L.N. Lipatov, Nucl. Phys. B213(1983)122

\bibitem{FK} M.~Fippel and R.~Kirschner, J. Phys. G17(1991)421 

\bibitem{g1} J.~Bartels, B.I.~Ermolaev and M.G.~Ryskin, Zeit. f. Physik
C70(1996)273, {\it ibid} C72(1996)627


\bibitem{BL-borel} 
I.I. Balitsky and L.N.Lipatov, Yad.~Fiz. 28(1978)1597; \\
I.I. Balitsky and L.N. Lipatov, Pisma v ZETF 30(1979)383


\bibitem{L-conf} I.I.~Balitsky, V.S.~Fadin and L.N.~Lipatov,
"Regge processes in non-Abelian gauge theories", in Proceedings
of XIV LNPI Winter School (Leningrad, 1979) p. 109

\bibitem{BFKL}
E.A.Kuraev, L.N.Lipatov and V.S.Fadin:
  Phys. Lett. 60B(1975)5;
Sov. JETP  71(1976)840;  Sov. JETP  72(1977)377 \\
Ya.Ya Balitsky and L.N.Lipatov, Yad.~Fiz. 28(1978)1597





\bibitem{impactf}
H.H.~Cheng and T.T.~Wu, Phys. Rev. 182(1969)1852, \\
G.V.~Frolov and L.N. Lipatov, Yad. Fiz. 13(1971)588



\end{thebibliography}
\end{document}